\documentclass[a4paper]{article}

\usepackage{fixltx2e}
\usepackage{graphicx}
\usepackage[hidelinks]{hyperref}
\usepackage[numbers,sort&compress]{natbib}
\usepackage{setspace}
\usepackage[autostyle=false, style=english]{csquotes}
\usepackage{float}
\usepackage{booktabs}
\usepackage{tabularx}
\usepackage{fancyvrb}
\usepackage{algpseudocode,algorithm}
\usepackage{caption}

\usepackage[table,xcdraw,dvipsnames]{xcolor}
\usepackage{xspace}
\usepackage{listings}
\usepackage{lstlinebgrd}
\usepackage{subcaption}
\usepackage{amsfonts}
\usepackage{amsthm,amsmath,amssymb,latexsym}
\usepackage[capitalize,noabbrev]{cleveref}
\usepackage{adjustbox}
\usepackage{algorithm}
\usepackage{algpseudocode}
\usepackage{multicol}

\usepackage{tex-custom/custom-listings}

\MakeOuterQuote{"}

\MakeRobust{\Call}

\usepackage{pifont}
\newcommand{\cmark}{\LARGE\textcolor{PineGreen}{\ding{51}}}%
\newcommand{\xmark}{\LARGE\textcolor{Red}{\ding{55}}}%

\Crefname{lstlisting}{Listing}{Listings}

\floatstyle{plain}
\restylefloat{figure}
\restylefloat{table}

\newcommand{\ie}{\textit{i.e.,}\xspace}

\newcommand{\souffle}{Souffl{\'e}\xspace}

\newcommand{\evmcode}[1]{{\textsc{#1}}}%

\newcommand{\eADD}{\evmcode{add}\xspace}

\newcommand{\eORIGIN}{\evmcode{origin}\xspace}
\newcommand{\eCALLER}{\evmcode{caller}\xspace}

\newcommand{\eGAS}{\evmcode{gas}\xspace}
\newcommand{\eJUMPDEST}{\evmcode{jumpdest}\xspace}

\newcommand{\eCALL}{\evmcode{call}\xspace}

\newcommand{\eSELFDESTRUCT}{\evmcode{selfdestruct}\xspace}

\begin{document}

\title{Vandal: A Scalable Security Analysis Framework for Smart Contracts}

\def \usyd {\small The University of Sydney}
\def \aspace {32pt}

\newcommand{\specialcell}[2][c]{%
  \begin{tabular}[#1]{@{}c@{}}#2\end{tabular}}
\newcommand{\email}{\small\texttt}
\date{}

\author{%
{\setstretch{0.8}\large
\resizebox{\columnwidth}{!}{%
\begin{tabular}{ccc}
  \multicolumn{1}{c}{\specialcell{%
      { Lexi~Brent%
      \textsuperscript{*}\vphantom{\footnote{Joint first authorship.}}}\\
    \usyd \\
    \email{lexi.brent@sydney.edu.au}
  }} &
  \multicolumn{1}{c}{\specialcell{%
    {
    Anton~Jurisevic%
    \textsuperscript{*}}\\
    \usyd\\
    \email{ajur4521@uni.sydney.edu.au}
  }} &
  \multicolumn{1}{c}{\specialcell{%
    {
    Michael~Kong%
    \textsuperscript{*}}\\
    \usyd\\
    \email{mkon1090@uni.sydney.edu.au}
}} \\ [\aspace]
  \multicolumn{1}{c}{\specialcell{%
    {
    Eric~Liu%
    \textsuperscript{*}}\\
    \usyd\\
    \email{eliu9480@uni.sydney.edu.au}
}} &
  \multicolumn{1}{c}{\specialcell{%
    {
    Fran\c{c}ois~Gauthier}\\
    \small Oracle Labs, Australia\\
    \email{francois.gauthier@oracle.com}
}} &
  \multicolumn{1}{c}{\specialcell{%
    {
    Vincent~Gramoli}\\
    \usyd\\
    \email{vincent.gramoli@sydney.edu.au}
}} \\ [\aspace]
\multicolumn{3}{c}{%
  \begin{tabular}{cc}
  \specialcell{%
     {
     Ralph~Holz}\\
     \usyd\\
     \email{ralph.holz@sydney.edu.au}
  } &
  \specialcell{%
    {
    Bernhard~Scholz}\\
    \usyd\\
    \email{bernhard.scholz@sydney.edu.au}
  }
\end{tabular}
}
\end{tabular}
}
}
}

\maketitle

\begin{abstract}
The rise of modern blockchains has facilitated the emergence of smart
contracts: autonomous programs that live and run on the blockchain.  Smart
contracts have seen a rapid climb to prominence, with 
applications predicted in law, business, commerce, and governance.

Smart contracts are commonly written in a high-level language such as
Ethereum's Solidity, and translated to compact low-level bytecode for
deployment on the blockchain. Once deployed, the bytecode is autonomously
executed, usually by a 
virtual machine. As with all
programs, smart contracts can be highly vulnerable to malicious attacks due
to deficient programming methodologies, languages, and toolchains,
including buggy compilers. At the same time, smart contracts are also
high-value targets, often commanding large amounts of cryptocurrency.
Hence, developers and auditors need security frameworks capable of
analysing low-level bytecode to detect potential security vulnerabilities.

In this paper, we present Vandal: a security analysis framework for Ethereum
smart contracts.  Vandal consists of an analysis pipeline that converts
low-level Ethereum Virtual Machine (EVM) bytecode to semantic logic
relations. Users of the framework can express security analyses in a
declarative fashion: a security analysis is expressed in a logic
specification written in the \souffle language. 
We conduct a large-scale empirical study for a set of common smart contract security 
vulnerabilities,
and show the effectiveness and 
efficiency of Vandal.  Vandal is both fast and
robust, successfully analysing over 95\% of all 141k unique contracts with
an average runtime of 4.15 seconds; outperforming the current state of the
art tools---Oyente, EthIR, Mythril, and Rattle---under equivalent
conditions.
\end{abstract}

\maketitle

\section{Introduction}\label{sec:intro}

Since the introduction of the Bitcoin cryptocurrency in 2008~\cite{bitcoin},
blockchain technology has seen growing interest from economists, lawyers, the
technology industry and
governments~\cite{Tapscott2016,Corda2017,Flood2015,UK2016}.  Blockchains are
decentralized distributed public ledgers, and have recently been used as
Turing-complete computational devices for storing and executing autonomous
programs called smart contracts.  Ethereum~\cite{eth-whitepaper} and
Cardano~\cite{cardano} are two examples of blockchains with smart contract
capabilities. Ethereum has become the \textit{de facto} standard platform for
smart contract development, with a market capitalization of \$20B
USD~\cite{ethcap}. For this reason, we focus exclusively on Ethereum smart
contracts.

Smart contracts are typically written in a high-level language such as
Solidity~\cite{solidity}, compiled to a low-level bytecode, and deployed on the
blockchain. Smart-contracts have unique addresses that are used to identify
them on the block-chain. Smart-contracts can then be invoked by users of the network or other smart
contracts, and are executed by the Ethereum Virtual Machine (EVM). Each smart
contract commands its own balance of Ether, the cryptocurrency used in
Ethereum.

Once deployed on the blockchain, a contract's bytecode becomes immutable. This
is a high-risk, high-stakes paradigm: deployed code is impossible to patch, and
contracts collectively control billions of USD worth of Ether.
As a consequence, security bugs in smart contracts can have disastrous
consequences.
A well-known example is the 2016 attack on a smart contract known as the ``The
DAO'', where an attacker exploited a reentrancy vulnerability, gaining control
of 3.6M Ether, worth more than \$50M USD at the
time~\cite{daohack,Leising2017}. 
One year later, in July 2017, an attacker exploited a vulnerability in a
library contract used by the "Parity multisig wallet", stealing 150k Ether
worth \$30M USD~\cite{ParityHack1,BlockCAT:Multisig,Qureshi2017}. Later that
year, in November 2017, an attacker exploited another vulnerability in a newer
version of the Parity multisig wallet contract, permanently freezing over 500k
Ether worth \$155M USD~\cite{ParityHack2}.
In January 2018, an attacker exploited an integer overflow vulnerability in the
contract underpinning the "Proof of Weak Hands" contract, making
off with 866 Ether worth \$2.3M USD~\cite{powhhack}. 

A wide range of known security vulnerabilities have been
described~\cite{Atzei2016,consensys2018,kalra2018,Luu2016,Nikolic2018,txorigin,chen2017}.
So-called \textit{unchecked send} vulnerabilities arise when the success/failure return
value of a message call operation is not checked, \ie always assuming
success. This vulnerability is one example of a larger class of
vulnerabilities arising from mishandled exceptions, and occurs surprisingly
frequently~\cite{Atzei2016,scanuncheckedsend16,consensys2018}.
\textit{Reentrancy} vulnerabilities emerge when a contract is not programmed
with reentrancy in mind, allowing an attacker to make reentrant message calls
that exploit an intermediate state~\cite{Atzei2016,consensys2018,Luu2016}.
\textit{Unsecured balance} vulnerabilities occur when a contract's balance is
exposed to theft by an arbitrary caller~\cite{Nikolic2018}. This may be due
to programmer error; for example, a misnamed constructor function that
inadvertently becomes a public function~\cite{Atzei2016}.
\textit{Destroyable contracts} are those which contain a \eSELFDESTRUCT instruction
on an exposed program path, \ie without adequate authentication, allowing an
arbitrary caller to permanently destroy the contract~\cite{Nikolic2018,Atzei2016}.
\textit{Origin vulnerabilities} occur when a contract performs authentication
by checking the return value of \eORIGIN rather than
\eCALLER~\cite{consensys2018,txorigin}.
In EVM, \eORIGIN returns the address
of the account that initiated the transaction, whereas \eCALLER returns the
address of the account or contract that initiated the currently executing
message call. An \eORIGIN vulnerability can be exploited if control is passed
to a malicious contract that makes a message call to the vulnerable contract:
the vulnerable contract checks \eORIGIN and sees the transaction initiator's
address, rather than the malicious contract's address.
Other vulnerability classes include those related to gas
consumption~\cite{Atzei2016,chen2017}, locking away of
Ether~\cite{Nikolic2018}, making calls to dead or unknown
contracts~\cite{Nikolic2018,Atzei2016}, timestamp
dependence~\cite{kalra2018,Luu2016}, transaction-ordering
dependence~\cite{kalra2018,Luu2016}, integer
overflow/underflow~\cite{kalra2018}, block state dependence~\cite{kalra2018},
Ether lost in transfer~\cite{Atzei2016}, and EVM's stack size
limit~\cite{Atzei2016}.

Security vulnerabilities in smart contracts stem from a wide range of issues
including programmer error, language design issues, and toolchain bugs such as
those in the Solidity compiler~\cite{solcbugs}.

\begin{figure}[tb]
\begin{center}
\includegraphics[width=0.6\textwidth]{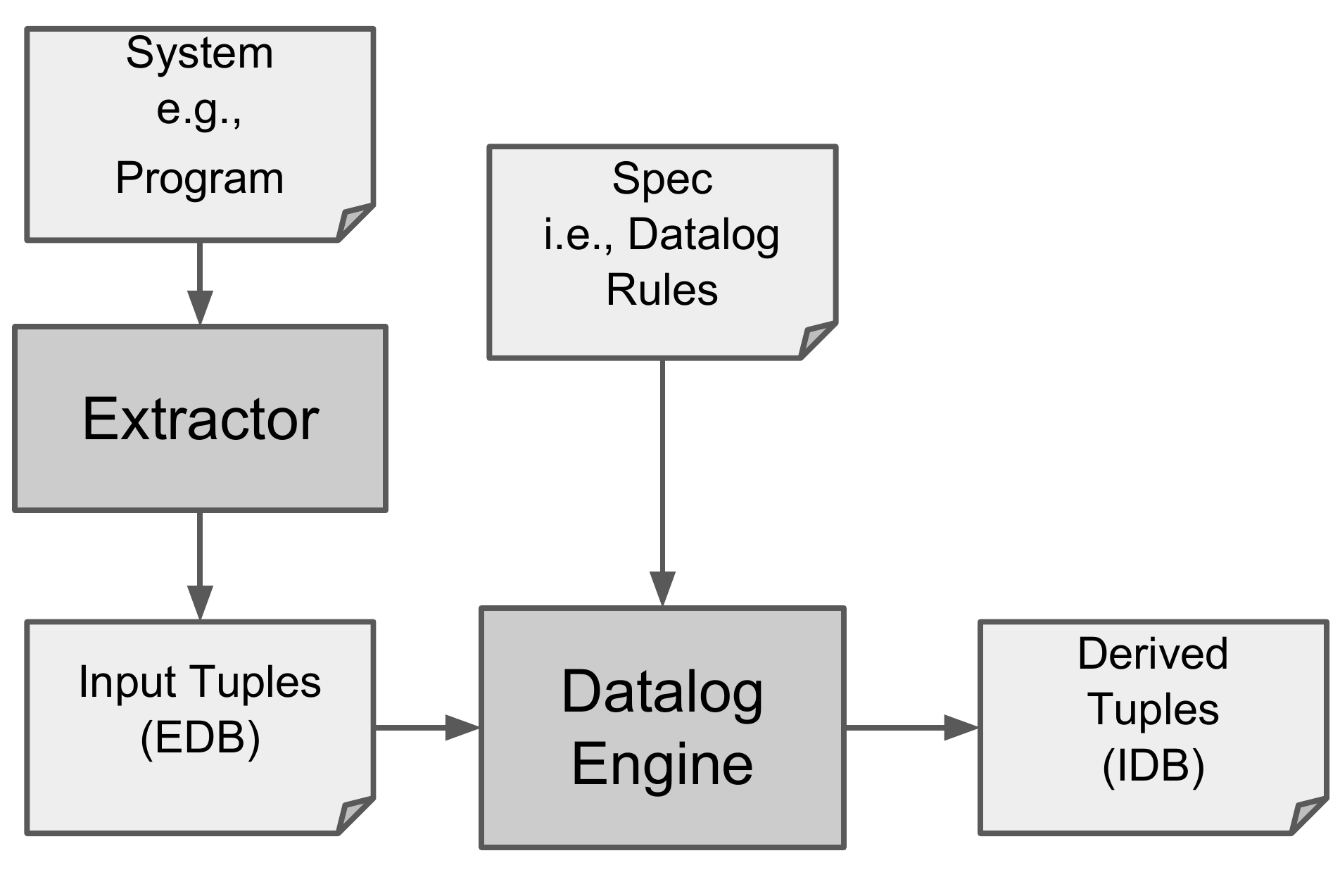}
\end{center}

\caption{Logic-Driven Program Analysis Approach: a program is converted by an
``extractor'' to a relational format also known as an Extensional Database
(EDB). The logic rules express the program analysis. A Datalog engine computes
the result of the program analysis from the EDB and the set of rules.  This
result is also known as an Intensional Database (IDB), and contains
intermediate and final results of the analysis.}
\label{fig:logic-driven}
\end{figure}

To analyze and alleviate the current deficiencies in smart contract programming
methodologies, language design, and toolchains, we present Vandal: a new
security analysis framework capable of directly analyzing smart contract
bytecode\footnote{Available online: \url{https://github.com/usyd-blockchain/vandal}}. 
Vandal facilitates a logic-driven static
program analysis approach~(cf.~Chapter 12.3 of \cite{Aho07},  and
\cite{whaley2005}) as depicted in~\cref{fig:logic-driven}.
In this approach, Datalog is used as a domain-specific language to bridge the
gap between program semantics of security vulnerabilities and the
implementation of the corresponding analyses.

In Vandal, security analysis problems are specified declaratively using
logic rules. 
An off-the-shelf Datalog engine then executes the specification for a set of
input relations that encode the contract (also known as the extensional
database) and produces an output relation containing a list of detected
security vulnerabilities and their locations in the bytecode.  Our
logic-driven approach results in security analyzers that are easy to write,
maintain, and less error-prone compared to low-level, hand-crafted
implementations. More importantly,
our approach allows users to explore the design space of security analyzers
without embarking on the difficult endeavour of writing/modifying a
hand-crafted static analyzer. This paradigm is supported by a cornucopia of
state-of-the-art Datalog engines that specifically target static
program analysis~\cite{whaley2005,Hoder.BM.11-muZ,logicblox}.

Vandal consists of two parts. First, an analysis pipeline translates
low-level bytecode to logic relations for the logic-driven security analysis.
This pipeline is represented by the ``extractor'' component
of~\cref{fig:logic-driven}. In Vandal, the logic relations expose data- and
control-flow dependencies of the bytecode. The second component is a set of
logic specifications for security analysis problems. Vandal uses
\souffle~\cite{jordan2016} as a Datalog engine, which synthesizes highly
performant C++ code from logic specifications. 

The contributions of our work are:
\begin{itemize}
   \item The Vandal security analysis framework that transforms low-level EVM
     bytecode to logic relations, enabling a logic-driven approach for
     expressing security analyzers. Vandal's analysis pipeline consists of a
     bytecode scraper that retrieves EVM bytecode from the blockchain, a
     disassembler that translates bytecode into opcodes, a decompiler that
     translates low-level bytecode to register transfer language, and an extractor
     that translates this register transfer language into logic semantic relations. 
   \item A new decompilation technique for incrementally reconstructing control-flow, 
         by applying symbolic execution for basic-blocks, 
        incremental data-flow analysis, and
        node-splitting techniques to disambiguate jump targets. 
        The precise detection of jump targets is essential for identifying stack locations  statically so 
         that the stack location can be assigned a register.
   \item A static analysis library containing logic specifications that expose
     control-flow graph properties, domain-specific properties of EVM
     operations, and data and control dependencies.
   \item A case study in which we phrase common security analyses for
     smart contracts as logic specifications within Vandal. We implement
     analyses for unchecked send, reentrancy, unsecured balance,
     destroyable contract, and use of \eORIGIN vulnerabilities in Vandal.
   \item A large-scale empirical analysis of all 141k unique smart contracts
     scraped from the public blockchain. We demonstrate that Vandal is more
     robust and efficient than Oyente~\cite{Luu2016}, EthIR~\cite{Albert2018},
     Mythril~\cite{HITBSecConf18}, and Rattle~\cite{rattle}, successfully
     decompiling over 95\% of all contracts with an average runtime of 4.15
     seconds.
\end{itemize}

The organization of this technical report is as follows:
In~\cref{sec:framework}, we introduce the stages of the Vandal analysis
pipeline and their implementation.
In~\cref{sec:specs}, we provide a use-case study showing how security analyses
for common vulnerabilities can be implemented in Vandal as logic
specifications. 
In \cref{sec:experiments}, we present and discuss results from our empirical
analysis experiments.
Related work is surveyed in \cref{sec:relatedwork}. 
Finally, in \cref{sec:conclusion} we draw our conclusions.

\section{The Vandal Framework}\label{sec:framework}

Vandal has been designed for phrasing security vulnerability analyses in a
declarative language called \souffle~\cite{jordan2016}. Expressing
vulnerability analyses in a declarative language allows security experts to
rapidly prototype new analyses and compose existing analyses with
ease~\cite{Bravenboer09,Aho07,whaley2005}. The logic-specification 
driven approach has several advantages, including this agile capability 
for designing and implementing security analyses. 
The design space for security analyses can be explored in terms of a precision
vs.\ time trade-off.
A traditional implementation of a security analysis would require several
hundreds of thousands of lines of code, becoming very costly to implement,
test, and maintain.
The performance penalties incurred by the usage of a logic language for
expressing security analyses is alleviated by the presence of modern logic
synthesizers such as \souffle that produce highly performant C++ code from
logic specifications. The produced C++ code is equivalent or better in
performance than hand-written code for program analysis~\cite{jordan2016}.

The challenge in the design of Vandal is the translation of smart contracts
into logic relations. The Vandal framework utilizes an analysis pipeline to
convert Ethereum bytecode stored on the blockchain to logic relations that
reflect the semantics of the smart contract. This is challenging: EVM bytecode
is executed by a very low-level stack-based abstract machine.  In contrast to
Java virtual machine bytecode, low-level EVM bytecode has no abstractions for
classes and methods,  no memory management,  no type checking,  no class
loading, and no notion of stack frames for function calls. The EVM is a very
simplistic computational device created for efficient storage of smart
contracts on a blockchain.  The control- and data-flows of a smart contract
are obfuscated by the virtual machine stack, making the task of static program
analysis immensely difficult.  To apply a logic-driven approach for analyzing
smart contracts, the EVM bytecode must be transformed to a new program
representation that reconstructs the data- and control- flow dependencies of
the original high-level language. 
    
\begin{figure}[tbp]
\includegraphics[width=\textwidth]{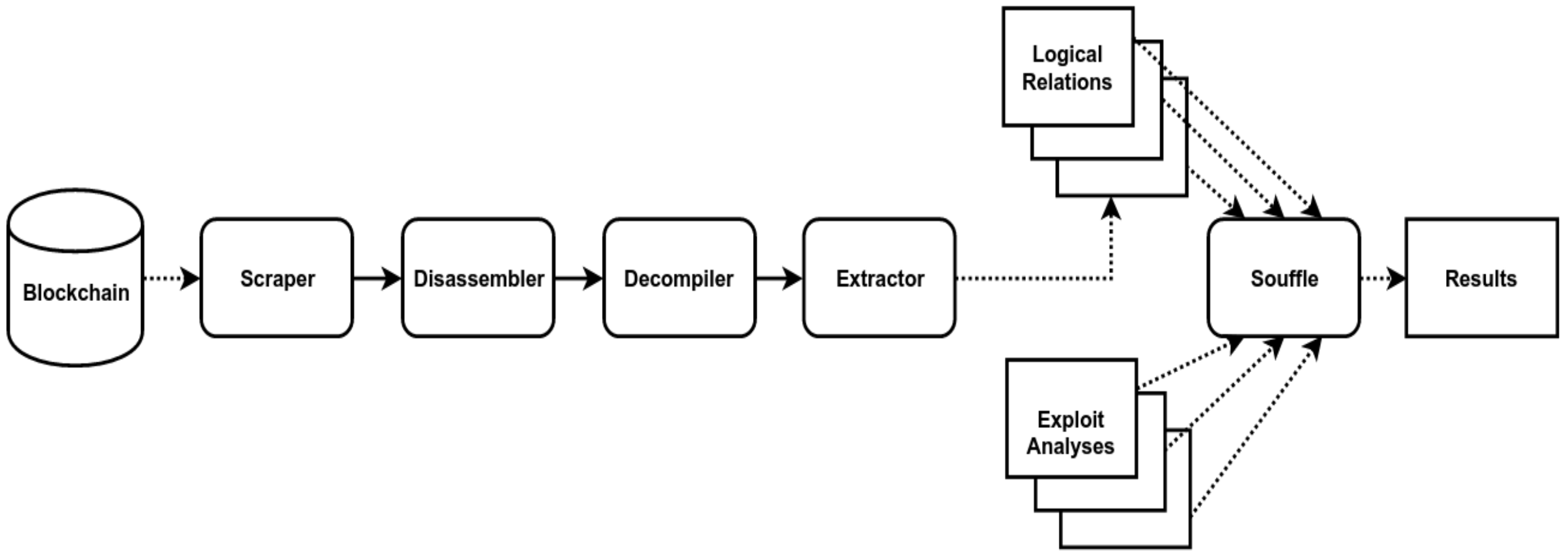}
\caption{Analysis Pipeline. 
The \textbf{Scraper:} extracts smart contract bytecode in bulk from the blockchain. 
The \textbf{Disassembler} converts the bytecode into mnemonics. 
The \textbf{Decompiler} translates the stack-based bytecode to a register transfer language. 
The \textbf{Extractor} produces logic relations from the register transfer
  language, reflecting the program semantics of the smart contract. 
The security analyses that report potential security vulnerabilities in
  the smart contracts are written in \souffle~\cite{jordan2016}.
}
\label{fig:security-pipeline}
\end{figure}

For the reconstruction of program semantics from Ethereum's low-level bytecode,
we introduce an analysis pipeline as shown in~\cref{fig:security-pipeline}. 
This pipeline breaks up the task of translating bytecode to an analyzable 
form into several stages, as follows:
The Ethereum bytecode is scraped from the blockchain by the \emph{Scraper}, disassembled by the 
\emph{Disassembler},  then decompiled into a register transfer language by the \emph{Decompiler}.
The decompiler reconstructs the control- and data-flow of the high-level language of the smart contract.
After decompilation, the bytecode program is expressed as a register-transfer language.
The \emph{Extractor} converts the intermediate representation of the register-transfer language
to logic relations. 
These logic relations capture the semantics of the smart contracts, and are
stored in a simple tab-separated value format. 
\souffle synthesizes the security analysis to executable programs that read the 
logic relations generated by the Extractor, and perform the security analysis for detecting 
vulnerabilities. Potential security vulnerabilities are reported after 
the execution of the security analysis.

In the following subsections we describe the design and implementation of
each pipeline stage. Note that the decompiler requires substantial research 
and implementation effort due to the incremental discovery of jump targets.
It is an intertwine approach of jump discovery, node-splitting, and 
incremental data-flow analysis. With the exception of the security analyses
and program analysis library, Vandal's infrastructure is written in Python.
The Vandal framework is open source and has been published under the BSD license~\cite{Vandal}.

\subsection{Scraper}

The first stage of our analysis pipeline is the scraper, which extracts the
bytecode representation of smart contracts from the live Ethereum blockchain.

We implement the scraper using the JSON-RPC API of the Parity~\cite{parity}
Ethereum client. For scraping all transactions, including "internal"
transactions, the Parity client must be synchronized with command-line flags
that enable tracing and disable pruning, as shown in~\cref{lst:paritycmd}.
\begin{plaincode}[
  caption={Parity command-line flags to enable tracing and disable pruning},
  label={lst:paritycmd},
]
\end{plaincode}
The scraping procedure is outlined in~\cref{alg:scraper}.
The procedure traverses through all transactions in all blocks, searching
for contract creations. The following helper functions are assumed to exist:
\begin{itemize}
  \item \textsc{Get-Code}(contract) --- 
    retrieves bytecode stored at the specified $contract$ address from the Parity
        client.
  \item \textsc{Trace-Transaction}(transaction) --- 
    retrieves the execution trace of the specified $transaction$ from the Parity
    client.
  \item \textsc{Save-Contract}(transaction, contract, code) --- 
    writes the given $contract$'s $code$ to disk, along with $transaction$
    metadata.
\end{itemize}

 \begin{algorithm}[tb]
 \caption{Ethereum blockchain smart contract scraping procedure.}\label{alg:scraper}
 \begin{algorithmic}[1]
   \Require start block number $s$ and end block number $e$
   \Require function \Call{Get-Code}{contract} 
   \Require function \Call{Trace-Transaction}{transaction}
   \Require function \Call{Save-Contract}{transaction, contract, code} 
   \For{each block $b$ such that $s <= b < e$}
     \For{each transaction $t$ in block $b$}
       \If{$t$ creates a contract at address $c$} 
         \State {\textit{code}=\Call{Get-Code}{$c$}}
         \State \Call{Save-Contract}{$t$, $c$,  \textit{code}}
       \Else
         \State $\textit{traces} \gets \Call{Trace-Transaction}{t}$
         \For{each \textit{trace} in \textit{traces}}
           \If{\textit{trace} created a contract at address $c$}
             \State $\textit{code} \gets \Call{Get-Code}{c}$
             \State \Call{Save-Contract}{$t$, $c$, \textit{code}}
           \EndIf
         \EndFor
       \EndIf
     \EndFor
   \EndFor
 \end{algorithmic}
 \end{algorithm}

Scraping the whole blockchain is inefficient due to its size when
synchronized with full tracing enabled ($>$ 1.5TB), Parity's underlying
hash-based database, and the overheads imposed by JSON-RPC HTTP requests. 
Parity's use of a hash-based key-value store, combined with a sequential
scraping process, results in slow random disk reads.
Our scraper uses various implementation techniques to reduce the runtime from
several days to less than a day:

\begin{enumerate}
  \item parallelization --- splitting the blockchain into contiguous chunks
    and scraping each portion in a separate instance;  
  \item increasing the number of JSON-RPC threads used by Parity (using the
    \code|--jsonrpc-threads| and \code|--jsonrpc-server-threads| command-line
    flags); and
  \item performing our scrapes with all I/O via a main-memory ramdisk (using a
    machine with 512GB of main memory).
\end{enumerate}

For each contract on the blockchain, our scraper produces a file on the local
filesystem containing the machine code of the smart contract, i.e., a sequence
of bytes that requires disassembly and decompilation in the later stages of
our pipeline. Due to legal reasons, the scraper has not been released as open
source. We will query the legal status of the scraper at a later point in time
so that it can be added to Vandal's open source repository on
GitHub~\cite{Vandal}.

\subsection{Disassembler}

The second stage of the analysis pipeline is the disassembler, which 
converts EVM bytecode to a series of readable low-level mnemonics annotated
with program counter addresses. This conversion is performed by a single
linear scan over the bytecode, converting each instruction to its
corresponding mnemonic and incrementing a program counter for each instruction
and each inline operand.
The result is a sequence of program addresses, mnemonics and their arguments.

\begin{figure}[tb]
\begin{solidity}
contract Factorial {
  function fact(uint x) returns (uint y) {
    if (x == 0) {
      return 1;
    } else {
      return fact(x-1) * x;
    }
  }
}
\end{solidity}
\caption{Example: high-level Solidity code.}
\label{fig:solidity-example}
\end{figure}

Assume that a programmer deploys a Solidity smart contract as shown
in~\cref{fig:solidity-example} on the blockchain. Vandal's scraper will
capture the machine code representation of this smart contract in the form of EVM
bytecode, producing a file such as the one illustrated
in~\cref{fig:disassembler}(a).
The disassembler converts the machine code to readable  EVM bytecode as shown
in~\cref{fig:disassembler}(b).

 \begin{figure}[tb]
\centering
\noindent\begin{minipage}[t]{.23\textwidth}
\begin{plaincode}[
  numbers=left
]
60606040526000357
c0100000000000000
00000000000000000
00000000000000000
00000000900480631
93ddd2c1460375760
35565b005b6042600
4805050605a565b60
40518082151581526
02001915050604051
80910390f35b60006
00560006000505414
9050606b565b9056

\end{plaincode}
\subcaption{Bytecode}
\end{minipage}\hfill
\begin{minipage}[t]{.70\textwidth}
\begin{minipage}[t]{.45\textwidth}
\begin{plaincode}[
  numbers=left,
]
0x00 PUSH1  0x60
0x02 PUSH1  0x40
0x04 MSTORE 
0x05 PUSH1  0xe0
0x07 PUSH1  0x02
0x09 EXP    
0x0a PUSH1  0x00
0x0c CALLDATALOAD 
0x0d DIV    
0x0e PUSH4  0x193ddd2c
0x13 DUP2   
0x14 EQ     
0x15 PUSH1  0x1a
0x17 JUMPI  
\end{plaincode}
\end{minipage}\hfill
\begin{minipage}[t]{.45\textwidth}
\begin{plaincode}[firstnumber=15]
0x18 JUMPDEST 
0x19 STOP   
0x1a JUMPDEST 
0x1b PUSH1  0x00
0x1d SLOAD  
0x1e PUSH1  0x05
0x20 EQ     
0x21 PUSH1  0x60
0x23 SWAP1  
0x24 DUP2   
0x25 MSTORE 
0x26 PUSH1  0x20
0x28 SWAP1  
0x29 RETURN 
\end{plaincode}
\end{minipage}
\subcaption{Disassembled bytecode}
\end{minipage}
\caption{Disassembler: converts a stream of EVM bytecode to a list of addresses/mnemonic pairs.}
\label{fig:disassembler}
 \end{figure}

The Vandal disassembler produces a similar output format to the Ethereum
Foundation's official disassembler, but with support for basic block boundary
delineation. The disassembler and decompiler share a common bytecode parsing
implementation.
An interface for declaring new instructions is provided in case changes occur
to the EVM specification.

\subsection{Decompiler}

The next stage in the analysis pipeline is the decompiler, which 
translates the low-level EVM bytecode to a register transfer language.
This register transfer language exposes data- and control-flow 
structures of the bytecode. Conceptually the semantics of the 
newly defined language has a strong overlap with those of EVM,
i.e., the instructions of the EVM are still reflected 
in the register transfer language of the decompiler. However,
the notion of a stack is replaced by a set of registers, and all
instructions operate on registers.
An abstract syntax of the language is given in~\cref{lst:ilgrammar} below.

\begin{plaincode}[
  caption={Syntax of the intermediate register transfer language.},
  label={lst:ilgrammar}
]
<operation> ::= Register = <rhs> | Op <args> 
<rhs>       ::= Register | Constant | Op <args>
<args>       ::= (Register | Constant ) *
\end{plaincode}

The language has two types of operations: operations that are side-effect free or have
a side-effect.  An operation may manipulate 
values in registers, memory, or storage.  The right-hand side of an operation 
can be either a constant, another register, or an operation that requires zero
or more arguments. The left-hand side is a register, a memory/storage location 
if the operation has a side-effect. 
All stack-based operations of the EVM can be 
expressed in this register language.
Note that the the storage structure
persists on the blockchain, whereas memory exists only transiently at runtime.

\begin{figure}
\begin{multicols}{3}
\begin{plaincode}[frame=no, numbers=none]
0x0: V0 = 0x60
0x2: V1 = 0x40
0x4: M[0x40] = 0x60
0x5: V2 = 0x10...0
0x23: V3 = 0x0
0x25: V4 = CALLDATALOAD 0x0
0x26: V5 = DIV V4 0x10...0
0x27: V6 = 0xb95d228
0x2d: V7 = EQ V5 0xb95d228
0x2e: V8 = 0x33
0x30: JUMPI 0x33 V7

0x31: JUMPDEST 
0x32: STOP 

0x33: JUMPDEST 
0x34: V9 = 0x4a
0x36: V10 = 0x4
0x38: V11 = CALLDATALOAD 0x4

0x39: JUMPDEST 
0x3a: V121 = 0x0
0x3d: V13 = 0x0
0x3f: V14 = EQ 0x0 S0
0x40: V15 = ISZERO V14
0x41: V16 = 0x65
0x43: JUMPI 0x65 V15

0x45: V17 = 0x1
0x47: V18 = 0x60
0x49: JUMP 0x60

0x4a: JUMPDEST 
0x4b: V19 = 0x40
0x4e: V20 = M[0x40]
0x51: M[V20] = S0
0x52: V21 = M[0x40]
0x56: V22 = SUB V20 V21
0x57: V23 = 0x20
0x59: V24 = ADD 0x20 V22
0x5b: RETURN V21 V24

0x5c: JUMPDEST 
0x5d: V25 = MUL S0 S1

0x60: JUMPDEST 
0x64: JUMP {0x4a, 0x5c}

0x65: JUMPDEST 
0x67: V26 = 0x5c
0x69: V27 = 0x1
0x6c: V28 = SUB S1 0x1
0x6d: V29 = 0x39
0x6f: JUMP 0x39
\end{plaincode}
\end{multicols}
\caption{Example: register transfer language.}
\label{fig:vandal-rtl}
\end{figure}

For example, consider  the code in~\cref{fig:vandal-rtl} that was decompiled
from the EVM bytecode of~\cref{fig:disassembler}.  The first EVM instruction
\verb|0x00 PUSH1 0x60| pushes the constant value \verb|0x60| onto the EVM
stack, as shown in the example of~\cref{fig:disassembler}.
In~\cref{fig:vandal-rtl}, the decompiler replaces this EVM instruction with
the register transfer instruction \verb|0x0: V0 = 0x60|.  The output is
generated such that instruction opcodes generally correspond directly to EVM
instructions excluding stack operations. The instructions have a program
address followed  by a colon. Note that \verb|M[x]| represents a memory
location \verb|x|.  Registers are denoted \verb|Vn| where \verb|n| is a
number. Register values can be read from or written to, however, our register
transfer language does not permit indirect access to registers.  An operation
may have several arguments that are separated by spaces (e.g. \verb|EQ V5 0xb95d228|).  Due to limitations of the decompilation, we may not be able to
determine jump addresses uniquely and we use curly braces to denote sets of
possible jump addresses  --- e.g. \verb|JUMP {0x4a, 0x5c}| For the translation
work it is important to realize that the decompiler does not produce
\emph{executable} register transfer language code due to the indeterminism
caused by unknown jump addresses.  The main purpose of the translation is the
detection of security vulnerabilities and hence the indeterminism can be seen
as a precursor to the program analysis expressed in logic.

To achieve the translation, the decompiler assigns to stack positions to
registers.  However, the control-flow of a smart contract is obfuscated by
stack-based data-dependencies. Hence, the decompiler must comprehend the
contract's stack access patterns, necessitating a deep program analysis.

The control-flow of an EVM bytecode program is initially unknown, requiring an
incremental analysis to iteratively construct the flow and determine the stack
locations.  To get a handle on the problem, we incrementally build a control
flow graph (CFG)~\cite{Aho07} and statically propagate constant values for to
identify potential jump addresses.  Note that in EVM bytecode, all valid jumps
must jump to a \eJUMPDEST instruction.  Using this property, we can easily
slice the disassembled bytecode into basic blocks.
However, determining the destination of a jump operation is difficult
since the EVM jump instructions do not have 
explicit arguments, and instead pop their operands from EVM's stack at
runtime.
These dependencies may be distinctly non-local, in that a value may be used
only after following a number of jumps. So in order to resolve these
dependencies, jump destinations must have been resolved beforehand.  Yet these
destination values may themselves have been defined non-locally.  So the
bytecode, even when disassembled into a sequence of more-readable opcodes, has
very little structure.

We use two phases to build the control-flow structure:
\begin{itemize}
\item In the first phase, we determine the basic blocks of the smart contract
  and assume a symbolic stack with symbolic values.  We symbolically execute
  each basic block and de-stackify its operations. Some registers may point to
  stack locations whose values were produced by the prior basic blocks. Other
  registers may point to stack locations that were produced by the current
  basic block.  We introduce symbolic labels for stack locations that are used
  to express data-dependencies across basic blocks, and try to resolve them
  via registers in the second phase.  For example, an \eADD EVM instruction
  that is placed as the first instruction of a basic block would generate a
  new register transfer operation \code|V = ADD S0 S1|, where \code|S0| and
  \code|S1| are the top symbolic stack elements which were written by an
  ancestral block, and \code|V| is a new register holding the result of the
  addition, which is pushed to the symbolic stack.

\item The second phase builds the control-flow graph incrementally. It
  resolving symbolic stack locations from the first phase, and new jump
  addresses emerge as a side effect. These new jump addresses, in turn, may
  lead to further addresses being discovered during the next iteration. This
  second phase is expressed as a fixed-point algorithm that propagates
  constant values that may carry jump addresses.
\end{itemize}

\begin{figure}
  \centering
  \includegraphics[height=0.85\linewidth]{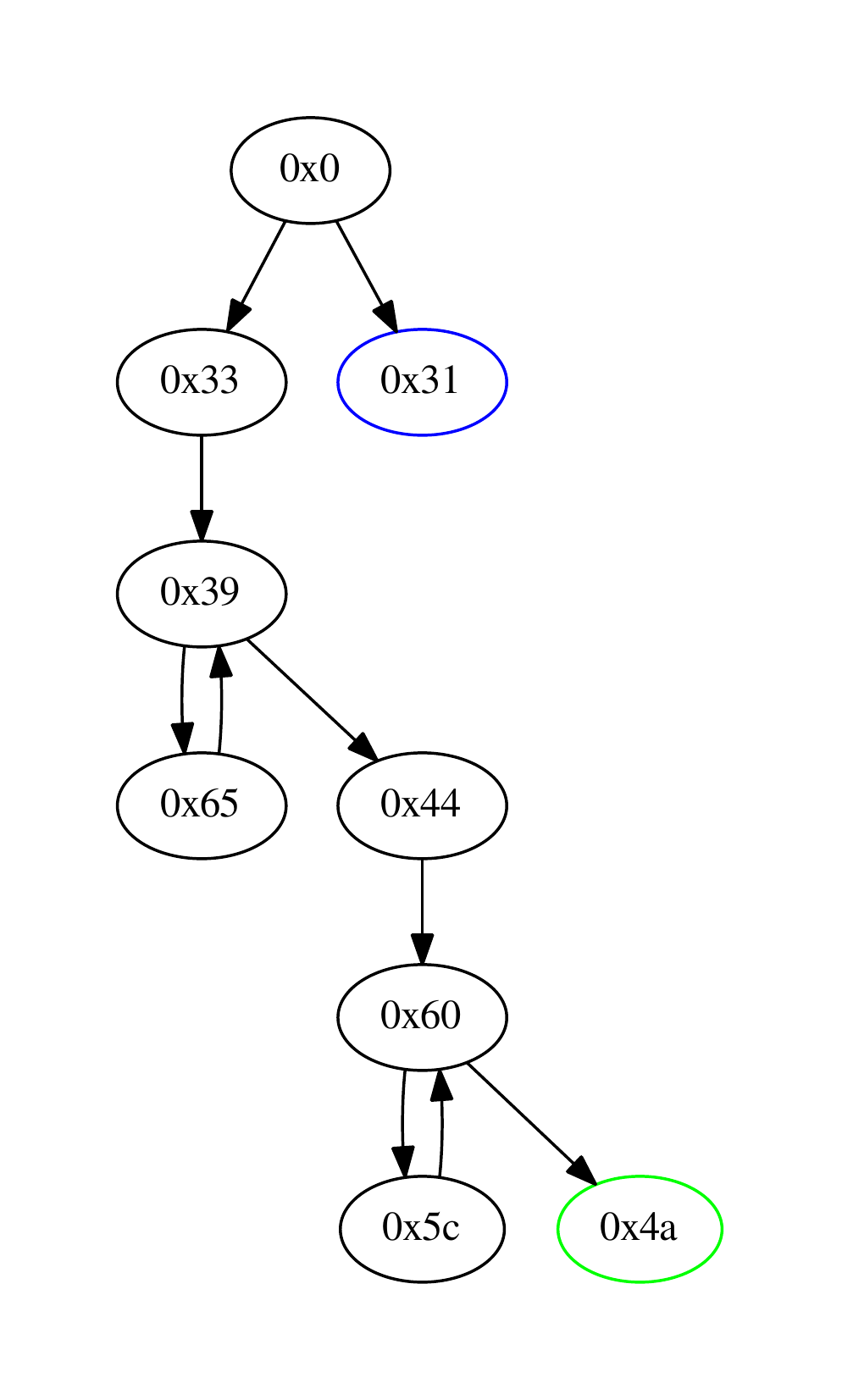}
  \captionof{figure}{Control flow graph for the \textsc{Factorial} contract,
  as output by Vandal.}
  \label{fig:vandalout}
\end{figure}

After decompilation, most jump addresses or sets of potential jump addresses
for basic blocks have been determined. For example, the corresponding control
flow graph of~\cref{fig:vandal-rtl} is shown in~\cref{fig:vandalout}.

\begin{figure}
\centering
\begin{subfigure}{.5\textwidth}
  \centering
  \includegraphics[height=\linewidth]{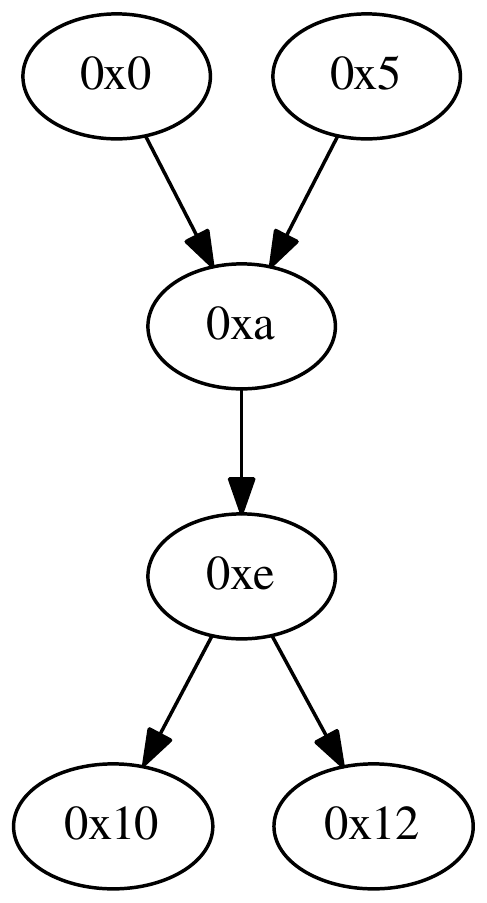}
  \caption{Before node splitting.}
  \label{fig:splitting-ex1}
\end{subfigure}%
\begin{subfigure}{.5\textwidth}
  \centering
  \includegraphics[height=\linewidth]{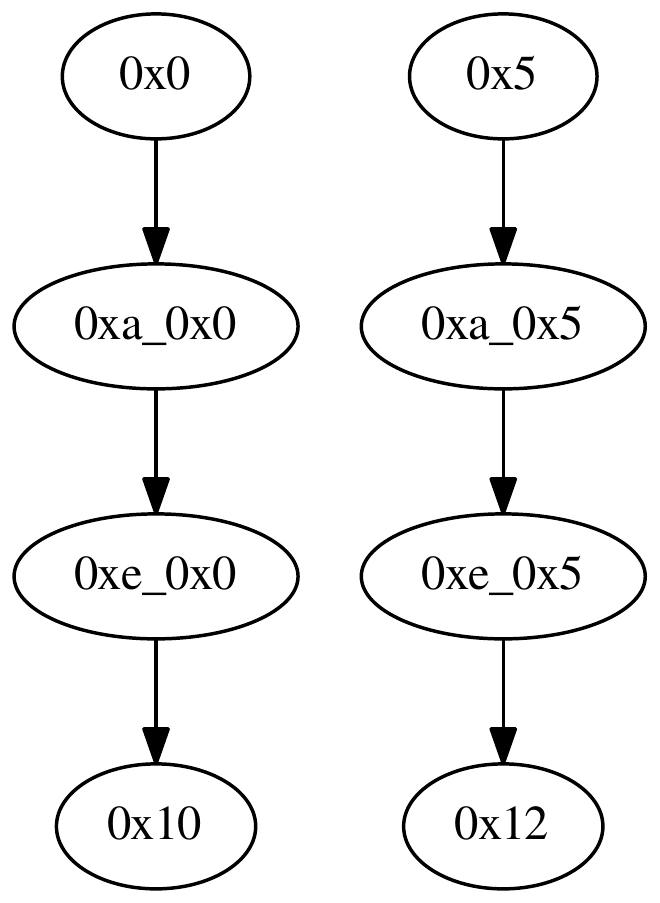}
  \caption{After splitting at node 0xE.}
  \label{fig:splitting-ex2}
\end{subfigure}
\caption{Demonstration of node splitting within Vandal.}
\label{fig:splitting}
\end{figure}

In order to obtain a precise dataflow analysis and resolve as many jump
destinations as possible, we implemented a node splitting technique in which
CFG nodes with multiple outgoing edges are split into several paths, each
originating at at a distinct common ancestor.
For example, consider the CFG in~\cref{fig:splitting-ex1}, and suppose that
the jump address used in block 0xE is pushed in 0xA's predecessors. In this
case, we have a set of two possible values \{0x10, 0x12\} --- one value from
each of 0xA's predecessors. However, if we split node 0xE by cloning the path
from 0xE to 0xA for each of 0xA's predecessors, we can resolve the jump
address at 0xE to a constant value for each path.  This information can then
be propagated down the graph during data flow analysis and used to determine
values in deeper nodes more precisely. The result of splitting at node
0xE is shown in~\cref{fig:splitting-ex2}.

After data flow analysis is complete, we merge duplicated nodes back together,
so that our final output is consistent with the input program, and does not
contain duplicated program counters, for instance.

\subsection{Extractor}

The next step in the analysis pipeline is the extractor that translates the register transfer language
to logic relations. Logic relations are comma-separated files that can be later read by the 
security analyses expressed in Souffl\'e~\cite{jordan2016}. The logic
relations express the register transfer language of an EVM bytecode program.

The logic relations use various domains such as the set of statements $S$, and the set of registers $V$ in 
a smart contract. There are other domains such as $O$ for the set of operations in the EVM. 
Edges of the control-flow graph are expressed by the relation $\textit{edge} \subseteq S \times S$.
The first statement in the control-flow graph is given by the singleton set $\textit{entry} \subseteq S$. 
For tracking data flow of smart-contracts, we introduce def-use sets~\cite{Aho07} that track
where registers are written to and read from, i.e., relation $\textit{def} \subseteq V \times S$ enumerates 
all write positions of a register, and relation $\textit{use} \subseteq V \times S$ enumerates all read positions of a register
in a smart-contract. The relation $\textit{op} \subseteq S \times O$ associates the op-code to a statement. 
The relation $\textit{value} \subseteq V \times {\mathbb N}$ associates constant values to registers 
if the set of potential constant values for the register is known.  The logic
relations are defined in \souffle as shown in~\cref{lst:logicrels}.

\begin{datalog}[
  caption={Definition of logic relations in \souffle.},
  label={lst:logicrels},float,floatplacement=H
]
.type Statement
.type Variable
.type Opcode
.type Value

.decl entry(s:Statement)                    
.decl edge(h:Statement, t:Statement)        
.decl def(var:Variable, stmt:Statement)   
.decl use(var:Variable, stmt:Statement, i:number)   
.decl op(stmt:Statement, op:Opcode)            
.decl value(var:Variable, val:Value)   
.input entry, edge, def, use, op, value
\end{datalog}

We have other relations that pre-compute dominance and post-dominance relations outside of Datalog
such that control-dependencies in smart contracts can be expressed in few lines of logic. We have specific 
logic relations for memory, storage, and jump operation to reflect the semantics of the smart contract
in more detail. These relations are required for the detection of some security analyses that
are presented in the next section. 

\section{Use-Case Study: Security Analyses in Vandal}\label{sec:specs}

To demonstrate the user experience for defining vulnerability analyses, we
describe some common vulnerabilities and how they can be implemented within
Vandal using Souffle Datalog code.
For simplicity, our explanations of each vulnerability use snippets of
Solidity source code.

\subsection{Vulnerability: Unchecked Send}

\paragraph{Explanation.} 
Solidity does not have exceptions that may be caught, so some low-level
operations report a success/failure status via a boolean return value.
One such operation is the \sol|address.send()| method, which performs a message call
and transfers Ether to a specified address.
If this operation fails, then any execution that occurred up until the point
of failure is rolled back, but \emph{execution of the calling function
continues as normal}. Hence, failing to check for and handle an error state
can lead to unintended effects, and is therefore considered to be dangerous
behaviour. 

Consider the Solidity function in \cref{lst:payfunc} which transfers 100 wei to
a creditor using \sol|creditor.send|, provided they have not already been paid:

\begin{figure}
  \centering
  \noindent\begin{minipage}[t]{.47\textwidth}
\begin{solidity}[
  caption={Vulnerable -- return value of send is not checked},
  label={lst:payfunc},
  linebackgroundcolor={\hly{3}},
  numbers=left,
]
function pay() {
  require(!paid);
  creditor.send(100);
  paid = true;
}
\end{solidity}
\end{minipage}\hfill
\begin{minipage}[t]{.47\textwidth}
\begin{solidity}[
  caption={Safe -- return value of send is checked and failure handled},
  label={lst:payfunc-fixed},
  numbers=left,
  linebackgroundcolor={\hlg{3}},
]
function pay() {
  require(!paid);
  require(creditor.send(100));
  paid = true;
}
\end{solidity}
\end{minipage}
\end{figure}

Assume that the account at the creditor address is itself a smart contract, and
its execution throws an exception. In this case, the funds will not be
transferred and the called function will fail. However, the state of the
contract is still updated as if the creditor had been paid as the next
operation in \sol|pay()| sets \sol|paid| to true. The state of the contract has
fallen out of sync with the reality that it is supposed to represent. The
creditor will no longer be able to obtain what they are owed as the funds are
now locked in the contract.

The correct implementation of this \sol|pay()| must check the return code of
\sol|creditor.send(100)|. If the return code indicates failure, it must not
update the contract state, e.g. by throwing an exception as shown in
~\cref{lst:payfunc-fixed}.

\paragraph{Implementation.} On the bytecode level, \sol|address.send|
corresponds directly to EVM's \eCALL instruction 
An analysis implementation for
unchecked send is shown in~\cref{lst:unchecked-dl}. Here, we check that the
return value \dlog|u| of a call operation (line 4) is neither used to control
the execution of a \sol|throw| (line 5), nor an update to persistent storage
(line 6). In other words, if no tangible action is taken based on the return
value of \eCALL, then the operation is flagged as vulnerable.
\begin{datalog}[
  caption={Analysis implementation in Vandal for unchecked send.},
  label={lst:unchecked-dl}
]
.decl uncheckedCall(u:Statement)

uncheckedCall(u) :- 
  callResult(_, u), 
  !checkedCallThrows(u), 
  !checkedCallStateUpdate(u).
\end{datalog}
The relations \dlog|callResult|, \dlog|checkedCallThrows|, and
\dlog|checkedCallStateUpdate| are all provided by Vandal's static analysis
library.

\subsection{Vulnerability: Reentrancy}
\label{sec:reentrancy}

\paragraph{Explanation.}
Another common mistake that can be much harder to spot is caused by
\textit{reentrancy}. When an Ethereum contract performs a message-call to
another contract by sending value or calling a function on the other contract,
the recipient contract may perform an arbitrary execution before returning
control to the caller. The recipient contract may, for example, call another
contract, or even the original contract. If the original contract is not
reentrancy-safe, i.e., guarantees that contract state is always correct
independent of reentering calls, then a malicious contract
can make reentrant calls that take advantage of intermediate state.
In 2016, a reentrancy vulnerability was infamously exploited by an unknown
attacker to "steal" then-equivalent of over \$50M USD from a contract known as
\textit{TheDAO}~\cite{daohack,Leising2017}. 

In~\cref{lst:withdraw-reent} we show an excerpt from a contract that stores
Ether for several users and keeps track of how much Ether is owned by each
user in an \sol|accounts| map. The contract allows users to withdraw their
share of Ether by calling \sol|withdraw()|, which reads the caller's balance
from \sol|accounts|, sends the Ether, and sets the new account balance to 0.
However, if the caller is a contract, it can make a reentrant call to
\sol|withdraw()| when control is passed to it by the message call on line 3.
At that point, the \sol|accounts| map has not yet been updated, so the
caller's balance would be sent again. Recursive reentrant calls of this nature
would allow an attacker to drain all Ether from the contract. Note, however,
that recursion is not a requirement: depending on the code, a single
reentrancy could be just as destructive.

\begin{figure}
\centering
\noindent\begin{minipage}{.47\textwidth}
\begin{solidity}[
  caption={Vulnerable -- \sol|accounts| is updated after message call},
  label={lst:withdraw-reent},
  linebackgroundcolor={\hly{3}},
  numbers=left
]
function withdraw() public {
  uint balance = accounts[msg.sender];
  msg.sender.call.value(balance)();
  accounts[msg.sender] = 0;
}
\end{solidity}
\end{minipage}\hfill
\begin{minipage}{.47\textwidth}
\begin{solidity}[
  caption={Safe -- all state is updated before message call},
  label={lst:withdraw-fixed},
  numbers=left,
  linebackgroundcolor={\hlg{3}},
]
function withdraw() public {
  uint balance = accounts[msg.sender];
  accounts[msg.sender] = 0;
  msg.sender.call.value(balance)();
}
\end{solidity}
\end{minipage}
\end{figure}

There is no way to prevent reentrant calls in Ethereum. Hence, all functions
must be reentrancy-safe. This can be achieved by using \sol|send()| instead of
\sol|call.value()()|, which does not forward enough gas for the callee to
perform computations. Alternatively, ensuring that all state changes occur
before any external calls would prevent an attacker from taking advantage of
intermediate state. Our revised function in~\cref{lst:withdraw-fixed}
demonstrates the latter approach, by simply swapping lines 3 and 4 of the
vulnerable code.

\begin{solidity}[
  caption={Example of a non-reentrant \eCALL protected by a mutex.},
  label={lst:mutexprotected},
  numbers=left,
]
function mutexProtected() {
  require(!mutex);
  mutex = true;
  <protected CALL code>
  mutex = false;
}
\end{solidity}

\paragraph{Implementation.}
At the bytecode level, we will consider any \eCALL operation to be reentrant
if it can be reached in a recursive call to the enclosing contract. This
simple definition can be generalised to a mapping from specific CALL
operations to program points they reach in a reentrant manner. We also require
that it forwards any remaining gas on to the callee, so that there is
sufficient gas for further execution to take place.  Finally, we require that
reentrant statements are not protected by a mutex-like structure. For
example,~\cref{lst:mutexprotected} is not reentrant, because the update of the
\texttt{mutex} field is carried through to recursive calls.

\begin{datalog}[
  caption={Analysis implementation in Vandal for reentrancy.},
  label={lst:reentrancy-dl}
]
.decl reentrantCall(stmt: Statement)

reentrantCall(stmt) :- 
  op(stmt, "CALL"), 
  !protectedByLoc(stmt, _), 
  gassy(stmt, gasVar), 
  op_CALL(stmt, gasVar, _, _, _, _, _, _).
\end{datalog}

An analysis for reentrancy vulnerabilities can be implemented in Vandal as
shown in~\cref{lst:reentrancy-dl}. Here, \dlog|protectedByLoc| is defined by
Vandal, and \dlog|protectedByLoc(stmt, _)| means that statement \dlog|stmt|
is protected by a mutex. The \dlog|gassy| relation is also defined by Vandal,
with \dlog|gassy(stmt, var)| implying that statement \dlog|stmt| uses a
variable \dlog|var| that depends on the result of a \eGAS operation.
In other words, a \eCALL is flagged as reentrant if it forwards sufficient gas
and is not protected by a mutex.

\subsection{Vulnerability: Unsecured Balance}

\paragraph{Explanation.}
Unsecured balance vulnerabilities arise when a contract's balance is left
exposed to theft. This can arise due to programmer error: for example, in
Solidity, a contract's constructor function is not qualified with a dedicated
keyword and giving it a name that differs from the contract's name turns it
into a public function.
Since it is a common pattern in contract constructors to set a state variable
to define the ``contract owner'', i.e., an address that can perform privileged
actions, a misnamed constructor often allows any caller to assume ownership
and access to privileged functionality.  Such a vulnerability was discovered
in the wild  in the ``Rubixi'' Ponzi scheme contract~\cite{Atzei2016}. The
Rubixi developers had renamed their contract at some point, but failed to
rename the constructor, allowing anyone to assume ownership and withdraw Ether
intended for the contract owner.

The contract in~\cref{lst:unsecbal} contains a misnamed constructor
called \sol|TaxOffice()|, which becomes a public function instead of a constructor. 
Anyone could call \sol|TaxOffice()| to take ownership of the contract and
then, by calling \sol|collectTaxes()|, withdraw Ether intended for the
contract's rightful owner.

\begin{figure}
\centering
  \noindent\begin{minipage}[t]{.47\textwidth}
\begin{solidity}[
  caption={Vulnerable -- misnamed constructor \sol|TaxOffice()| sets \sol|owner|},
  label={lst:unsecbal},
  linebackgroundcolor={\hly{4}},
  numbers=left
]
contract TaxMan {
  address private owner;
  ...
  function TaxOffice() {
    owner = msg.sender;
  }
  function collectTaxes() public {
    require(msg.sender == owner);
    owner.send(tax);
  }
}
\end{solidity}
\end{minipage}\hfill
\begin{minipage}[t]{.47\textwidth}
\begin{solidity}[
  caption={Safe -- \sol|owner| is initialised directly during contract creation},
  label={lst:unsecbal-fixed},
  numbers=left,
  linebackgroundcolor={\hlg{2}},
]
contract TaxMan {
  address private owner = msg.sender;
  ...
  function collectTaxes() public {
    require(msg.sender == owner);
    owner.send(tax);
  }
}
\end{solidity}
\end{minipage}
\end{figure}

In~\cref{lst:unsecbal-fixed}, the constructor function is eliminated
altogether in favour of direct initialization of the \sol|owner| state
variable.
Here, \sol|owner| is initialised when the contract is first created and is
never updated from within a function.

\paragraph{Implementation.} We say that a contract is vulnerable if an
arbitrary caller can force it to transfer Ether, and can manipulate the
address to which that Ether is transferred.

\begin{datalog}[
  caption={Analysis implementation in Vandal for unsecured balance.},
  label={lst:unsecbal-dl},float,floatplacement=H
]
.decl unsecuredValueSend(stmt:Statement)

unsecuredValueSend(stmt) :-
  op_CALL(stmt, _, target, val, _, _, _, _),
  nonConstManipulable(target),
  def(val, _),
  !value(val, "0x0"),
  !fromCallValue(val),
  !inaccessible(stmt).
\end{datalog}

Our example analysis implementation is shown in~\cref{lst:unsecbal-dl}. 
At the EVM level, Ether transfers are handled by the \eCALL instruction.
Hence, this specification checks for a \eCALL instruction that:
\begin{enumerate}
\item has a destination address that can be manipulated (line 5); and
\item transfers a nonzero amount of Ether (lines 5-7); and
\item does not simply forward the Ether from in the incoming message call
  (line 8); and
\item can be executed by an arbitrary caller (line 9).
\end{enumerate}
The relations \dlog|fromCallValue(var)| and \dlog|inaccessible(stmt)| are
provided by Vandal's static analysis library, with \dlog|inaccessible(stmt)|
meaning that the operation at \dlog|stmt| is either unreachable code, or its
execution is conditionally dependent on a value that can not be manipulated by
an arbitrary caller.

\subsection{Vulnerability: Destroyable Contract}

\paragraph{Explanation.}
An EVM instructions exists that is capable of nullifying the bytecode of a
deployed contract: \eSELFDESTRUCT. When called, it halts EVM execution,
deletes the contract's bytecode, and sends any remaining value to a designated
address.  In a pattern that is similar to exposed ownership control, this can
be exploited if this function is accessible for non-authorized callers.

\cref{lst:selfdestruct} shows such an accessible \eSELFDESTRUCT vulnerability.
The public function \sol|destroy()| should probably only be called by
authorized callers, but performs no authentication before calling
\sol|selfdestruct()|.

\begin{figure}
\centering
  \noindent\begin{minipage}[t]{.47\textwidth}
\begin{solidity}[
  caption={Vulnerable -- code path to \sol|selfdestruct()| is not protected},
  label={lst:selfdestruct},
  linebackgroundcolor={\hly{2}},
  numbers=left
]
function destroy() public {
  selfdestruct(msg.sender);
}
\end{solidity}
\end{minipage}\hfill
\begin{minipage}[t]{.47\textwidth}
\begin{solidity}[
  caption={Safe -- \sol|selfdestruct()| may only be executed by the contract owner},
  label={lst:selfdestruct-fixed},
  numbers=left,
  linebackgroundcolor={\hlg{2}},
]
function destroy() public {
  require(msg.sender == owner);
  selfdestruct(msg.sender);
}
\end{solidity}
\end{minipage}
\end{figure}

\cref{lst:selfdestruct-fixed} shows how easily this can be mitigated by
requiring that the caller of \sol|destroy()| be the contract's owner.

\paragraph{Implementation.}
We implement this in Vandal as shown in~\cref{lst:selfdes-dl} by requiring
that a \eSELFDESTRUCT instruction exists (line 5), and is directly reachable from
the contract's entry point with its execution not conditional upon a value
outside the attacker's control (line 4). The \dlog|inaccessible| relation is
defined by Vandal.

\begin{datalog}[
  caption={Analysis implementation in Vandal for accessible \eSELFDESTRUCT.},
  label={lst:selfdes-dl},
]
.decl destroyable(stmt:Statement)

destroyable(stmt) :- 
  !inaccessible(stmt), 
  op(stmt, "SELFDESTRUCT").
\end{datalog}

Any code path leading to a \eSELFDESTRUCT instruction, that can be executed by
an arbitrary caller, is almost certainly unintended behaviour and is flagged
by this implementation.

\subsection{Vulnerability: Use of \texttt{ORIGIN}}

\paragraph{Explanation.}

Solidity's low-level \sol|tx.origin| contains the address of the account that
made the \emph{first} message call in the currently executing transaction.
Hence, its use should be avoided when authenticating the sender of a message
call, otherwise a malicious intermediary contract could perform authenticated
message calls. Instead, programmers should use \sol|msg.sender|, which
contains the address that created the currently executing message call.

\begin{figure}
\centering
  \noindent\begin{minipage}[t]{.47\textwidth}
\begin{solidity}[
  caption={Vulnerable -- authentication mechanism checks \sol|tx.origin|.},
  label={lst:origin},
  linebackgroundcolor={\hly{2}},
  numbers=left
]
function protected() public {
  require(tx.origin == self.admin);
  // do something sensitive
}
\end{solidity}
\end{minipage}\hfill
\begin{minipage}[t]{.47\textwidth}
\begin{solidity}[
  caption={Safe -- authentication mechanism checks \sol|msg.sender|.},
  label={lst:origin-fixed},
  numbers=left,
  linebackgroundcolor={\hlg{2}},
]
function protected() public {
  require(msg.sender == self.admin);
  // do something sensitive
}
\end{solidity}
\end{minipage}
\end{figure}

A simple example of a broken authentication mechanism is shown
in~\cref{lst:origin}, with a correct version shown in~\cref{lst:origin-fixed}.

\paragraph{Implementation.}
In bytecode form, \sol|tx.origin| corresponds to EVM's \eORIGIN instruction,
and \sol|msg.sender| to \eCALLER. We can implement an analysis in Vandal by
checking that an \eORIGIN instruction exists, and that its result is used in
some other potentially sensitive operation such as a conditional or a write
to storage. This is demonstrated in~\cref{lst:origindl}.

\begin{datalog}[
  caption={Analysis implementation in Vandal for use of \eORIGIN.},
  label={lst:origindl},
]
.decl originUsed(stmt:Statement)

originUsed(stmt) :-
  op(stmt, "ORIGIN"),
  def(originVar, stmt),
  depends(useVar, originVar),
  usedInStateOrCond(useVar, _).
\end{datalog}

Here, \dlog|op| and \dlog|def| are relations output by Vandal's decompiler.
The \dlog|depends| relation is defined by Vandal, with \dlog|depends(x,y)|
meaning that the value in variable $x$ depends on the value in variable $y$.
The \dlog|usedInStateOrCond| relation is also defined by Vandal, with
\dlog|usedInStateOrCond(var, stmt)| meaning that variable $var$ is used in
$stmt$, and $stmt$ is an operation corresponding to a conditional or a write to
storage.

\section{Experimental Evaluation}\label{sec:experiments}

We perform an empirical evaluation of Vandal using a corpus consisting of all
141k unique smart contracts currently deployed on the live Ethereum
blockchain, in bytecode form.
We compare Vandal's performance to that of the Oyente~\cite{Luu2016},
EthIR~\cite{Albert2018}, Rattle~\cite{rattle}, and
Mythril~\cite{HITBSecConf18}, and show that Vandal outperforms all systems in
terms of average runtime.

\subsection{Bytecode Corpus}
We build our corpus by running Vandal's scraper (see~\cref{sec:framework}) on
the live Ethereum blockchain to retrieve bytecode for every contract deployed
as of block number 6237983 (2018-08-30).  In total there are 7.4M contracts,
of which only 141k have unique bytecode.

\begin{figure}
  \centering
  \includegraphics[width=0.8\columnwidth]{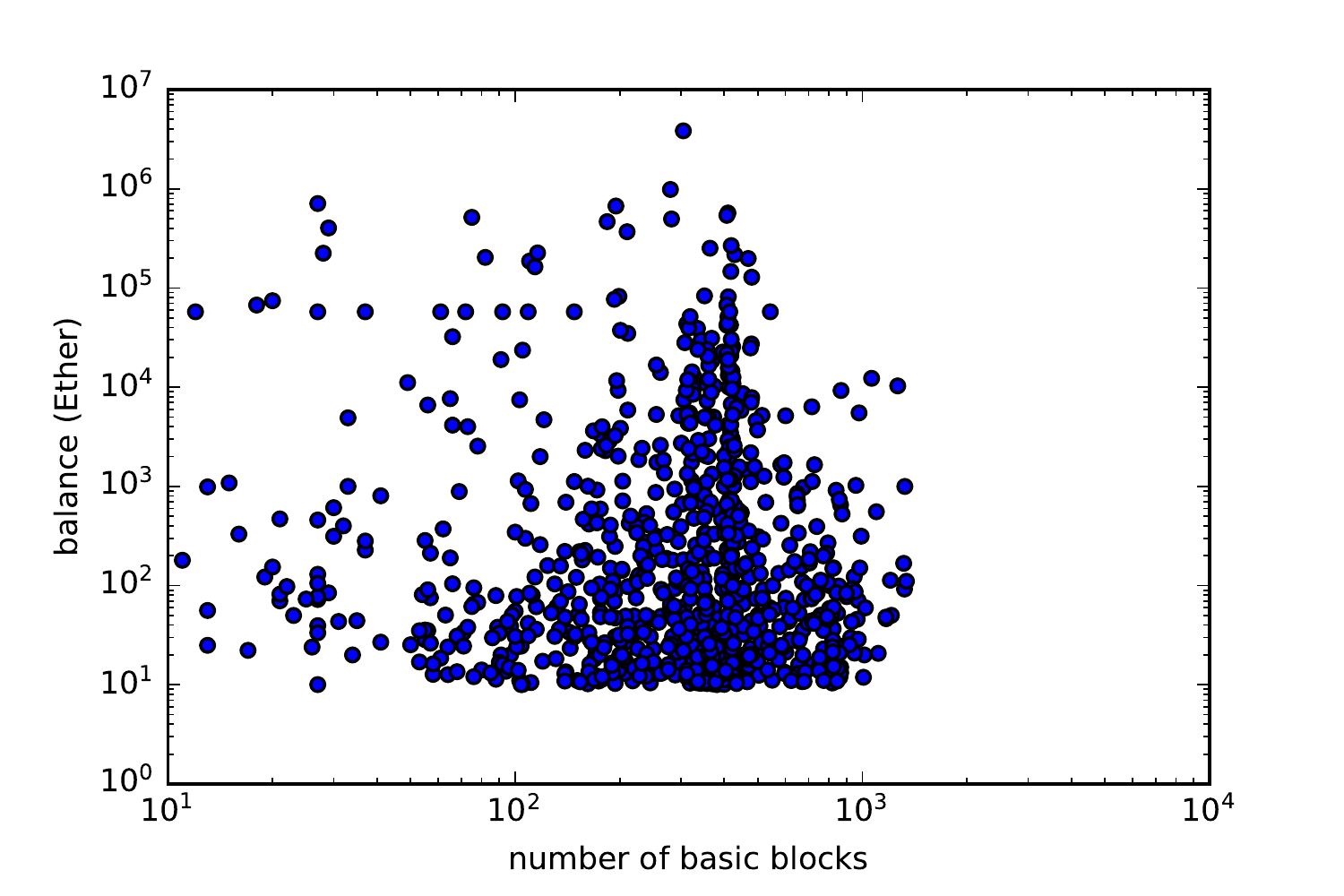}
  \caption{Number of basic blocks and Ether balance for all unique bytecodes.}
  \label{fig:blocksbal}
\end{figure}

\begin{figure}
  \centering
  \includegraphics[width=0.8\columnwidth]{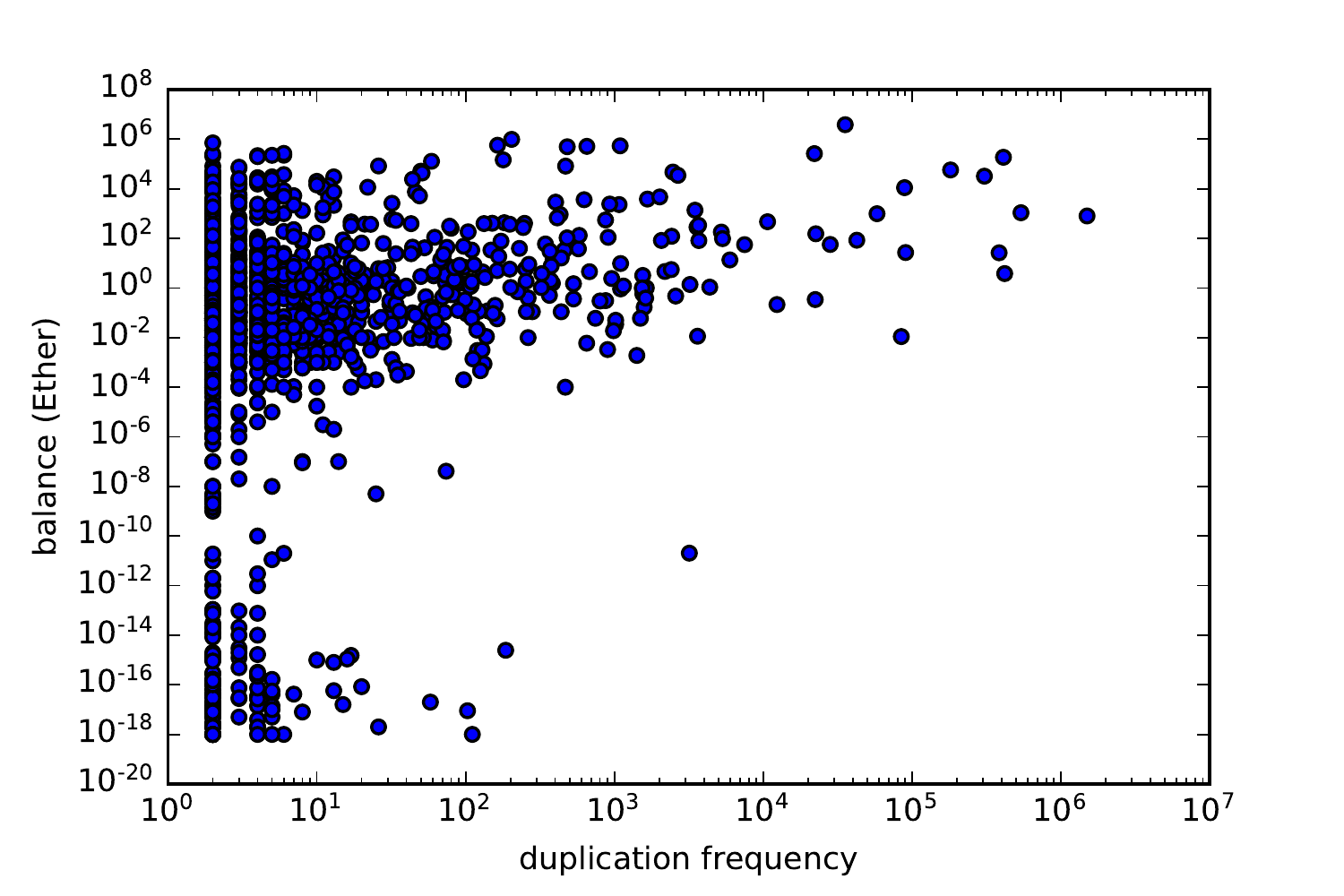}
  \caption{Duplication frequency and Ether balance for all unique bytecodes.}
  \label{fig:dupbal}
\end{figure}

The total amount of Ether controlled by each unique bytecode in our corpus is
shown in~\cref{fig:blocksbal}. We use the number of basic blocks as a simple
measure of complexity, and define `balance' to be the sum of the balances of
all contracts with the same bytecode.  We observe that bytecodes with a higher
complexity control greater amounts of Ether, and that most bytecodes are
non-trivial, consisting of between 100 and 1k basic blocks.  \cref{fig:dupbal}
shows all bytecodes that have a positive balance and occur more than once
among all 7.4M.  We observe that many heavily-duplicated contracts control
balances greater than 100 Ether, which is equivalent to \$29k USD as of this
writing.  These bytecodes are high-value targets: an exploit in just one of
these would allow an attacker to compromise and drain thousands of contract
accounts.

\subsection{Comparison of Tools}

We compare Vandal's performance to that of Oyente, EthIR, Rattle, and Mythril
using our corpus of 141k unique contract bytecodes. In our experiments, a
timeout of 60s is used for decompilation of each contract. This is necessary
to bound the overall runtime of the experiment. 
All experiments were run on a machine with an Intel Xeon E5-2680 v4 2.40GHz
CPU and 512GB of RAM. Other than our experimental workload, the machine was
idle. We used 56 concurrent analysis processes, since our machine has 56
logical cores.

\begin{table}
	\centering
  \caption{Runtime statistics for each tool for all successfully analyzed
  contracts. The total runtime includes both successful and unsuccessful
contracts.}
    \begin{tabularx}{\columnwidth}{XXXXX}
			\toprule
      Tool & Min (s) & Max (s) & Average (s) & Total (h) \\
      \midrule 
      Oyente & 0.26 & 60.00 & 13.68 & 9.69 \\
      EthIR & 0.25 & 59.99 & 11.99 & 8.89  \\
      Mythril & 1.47 & 59.99 & 11.10 & 13.51 \\
      Rattle & 0.12 & 60.00 & 4.47 & 3.56 \\
      Vandal & 0.29 & 59.99 & 4.15 & 8.08  \\
			\bottomrule
    \end{tabularx}
    \label{tab:runtimes}
\end{table}

As can be seen in~\cref{tab:runtimes}, Vandal outperforms all existing tools
in terms of average runtime per contract. The total runtimes shown here
represent total wall clock time, which is not correlated with the averages
because it includes contracts that time out or cause the tool to exit in an
error state.
We observe that the average runtime of all tools for successfully-decompiled
contracts is more than four times below our timeout threshold of 60s,
indicating that increasing the timeout further would likely result in
diminishing returns.

\begin{figure}
  \centering
  \includegraphics[width=\columnwidth]{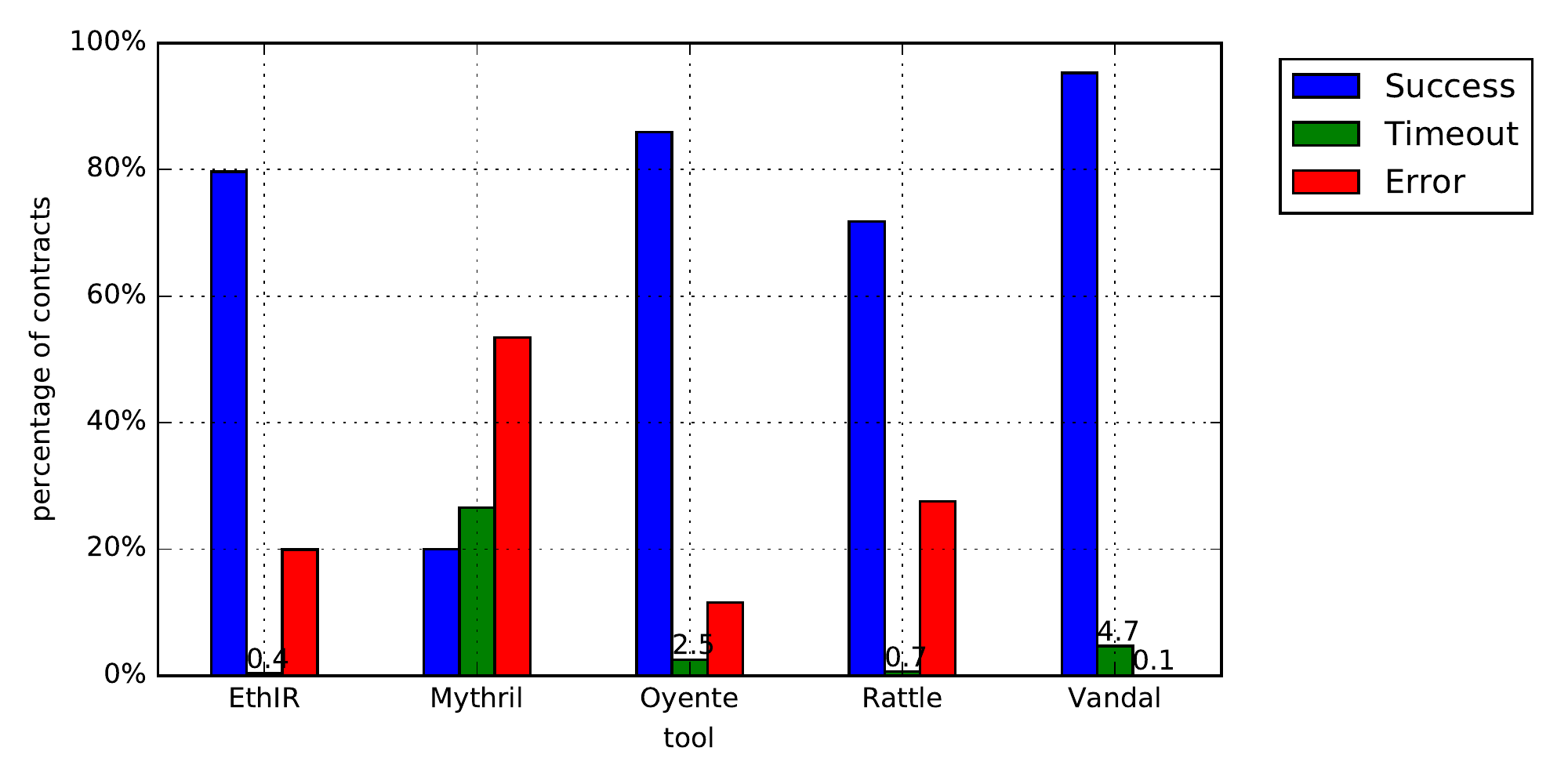}
  \caption{Percentage of contracts that resulted in successful decompilation,
  timeout, or error states with each analysis tool.}
  \label{fig:succ}
\end{figure}

In~\cref{fig:succ} we show the percentages contracts that were successfully
analyzed, vs.\ the percentages that resulted in an error or timed out after
60s, for each tool. 
We see that Vandal has the highest success rate and lowest error rate in
comparison to every other tool, successfully decompiling over 95\% of all
contracts and exiting in an error state for only 0.1\% of contracts.

\begin{table}
\centering
\caption{Comparison of tool support for our chosen vulnerability analyses.}
  \begin{tabularx}{\columnwidth}{Xccc}
    \toprule
    Vulnerability Analysis & Vandal & Mythril & Oyente \\
    \midrule
    Destroyable & \cmark & \cmark & \xmark \\
    Reentrancy & \cmark & \cmark & \cmark \\
    Unchecked Send & \cmark & \cmark & \xmark \\
    Unsecured Balance & \cmark & \cmark & \xmark \\
    Use of \eORIGIN & \cmark & \cmark & \xmark \\
    \bottomrule
  \end{tabularx}
\label{tab:compvuln}
\end{table}

\begin{figure}
  \centering
  \includegraphics[width=0.8\columnwidth]{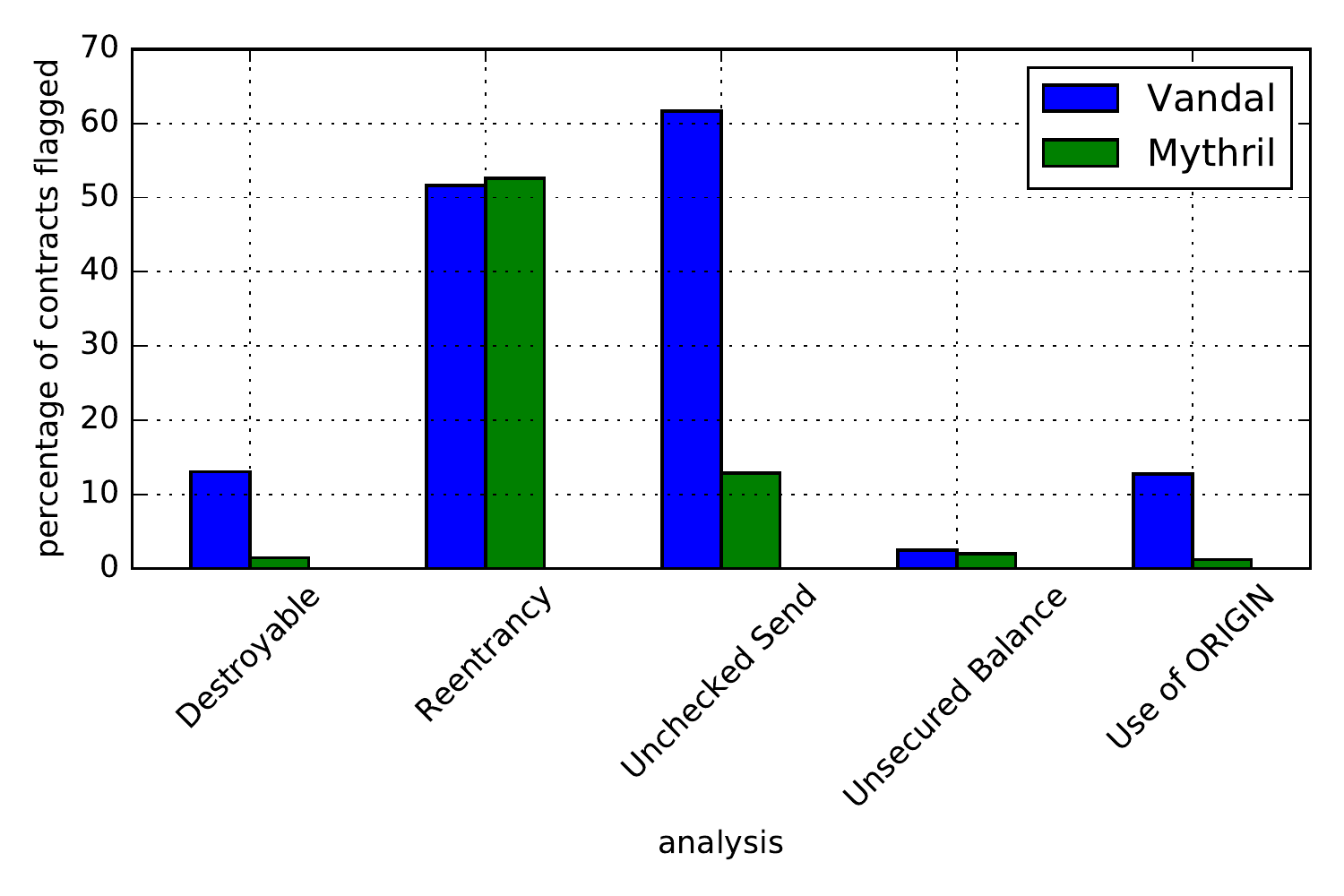}
  \caption{Percentage of contracts flagged as vulnerable for each analysis
  with Vandal and Mythril.}
  \label{fig:compvuln}
\end{figure}

Out of these tools, only Vandal, Oyente, and Mythril are capable of flagging
security vulnerabilities.  The detection capabilities of each tool with
respect to the analyses described in~\cref{sec:specs} are shown
in~\cref{tab:compvuln}.  (Note that Oyente and Mythril both implement analyses
for a wider range of vulnerability classes than just those shown
in~\cref{tab:compvuln}.)
In~\cref{fig:compvuln} we compare the percentage of successfully-decompiled
contracts that were flagged by Vandal and Mythril for each analysis.  With the
exception of reentrancy, we observe that Vandal flags a higher percentage of
contracts for all analyses.  This is expected: Vandal's use of abstract
interpretation ensures that \emph{all} possible executions are explored,
whereas Mythril's concolic analysis examines only a subset of execution
traces, possibly missing some true positives. For reentrancy, the difference
between Mythril and Vandal in~\cref{fig:compvuln} is less than 1\%.  We
attribute this discrepancy to noisy data, given the high error rate of Mythril
(see~\cref{fig:succ}).
Although Vandal flags more than 50\% of contracts for reentrancy and unchecked
send, some of the flagged contracts may not be practically exploitable, and
others may be false positives.

\section{Related Work}\label{sec:relatedwork}
Approaches and tools for analysis and verification of smart contracts have
received widespread attention in the literature due to the high-risk
environment of smart contract development.

\paragraph{Decompilers for Smart Contracts.}
Porosity \cite{porosity} is a high-level decompiler from EVM bytecode to
Solidity-like source that was introduced by M.~Suiche at DEF CON 25. 
Porosity is implemented in C++ and is intended to be a preliminary prototype for
decompiling EVM bytecode to high level Solidity source code.
The EthIR~\cite{Albert2018} framework is built upon the trace-based Oyente
tool~\cite{Luu2016} and performs high-level analysis of Ethereum
bytecode. It outputs an intermediate representation in which local variables
are introduced to each basic block, simplifying the analysis.
Although EthIR reconstructs fragments of high-level control- and data-flows
from Oyente traces, it does not improve on Oyente's control-flow graph
recovery capability.
Mythril is a security analysis tool for Ethereum smart
contracts~\cite{HITBSecConf18}.  Mythril performs decompilation aided by a
dynamic symbolic execution engine (Laser-EVM). It produces traces that are
used to generate an intermediate representation and it therefore suffers
similar incompleteness issues as EthIR.  Similarly, Rattle~\cite{rattle} also
constructs an IR in SSA form~\cite{Cytron:1991:ECS:115372.115320} and performs
program analysis on it.
The teEther~\cite{Krupp18} exploit generation tool, although not a decompiler,
operates on EVM bytecode and uses a form of iterative data-flow analysis to
reconstruct contract control flow graphs. teEther automatically generates
exploits for vulnerable contracts using symbolic execution with the Z3
constraint solver to solve path constraints.  It generates exploits by
focusing on "critical paths" in the control flow graph; those containing
sensitive state-changing operations such as inter-contract calls and writes to
persistent storage.  Unlike our Vandal framework's Datalog interface, however,
teEther does not have a focus on usability and rapid prototyping of new
vulnerability analyses.  teEther is implemented in Python and will be released
as open source, but is not yet available as of this writing.

\paragraph{Exploit Identification and Analysis.}
Previous works which applied static program analysis for smart contracts can
be classified according to their underlying techniques, including dynamic symbolic
execution, formal verification, and abstract interpretation. 

Systems including Oyente~\cite{Luu2016},
\textsc{maian}~\cite{Nikolic2018},
\textsc{gasper}~\cite{chen2017} and recent
work~\cite{Grossman2017} use an approach based on
dynamic \emph{symbolic execution} or trace semantics, which is fundamentally unsound
since only some program paths can be explored.

Semi-automated \emph{formal verification} approaches have also been
proposed~\cite{DBLP:journals/corr/abs-1802-08660,Hildenbrandt2017,hirai2017,Bhargavan2016,Why3,Amani2018}
for performing complete analyses of smart contracts using interactive theorem
provers such as Isabelle/HOL~\cite{Isabelle}, F*~\cite{Fstar},
Why3~\cite{Why3}, and $\mathbb K$~\cite{Kframework}. These approaches have a
common theme: a formal model of a smart contract is constructed and
mathematical properties are shown via the use of a semi-automated theorem
prover.  Recently, a complete small-step semantics of EVM bytecode has been
formalized for the F* proof
assistant~\cite{DBLP:journals/corr/abs-1802-08660}. Other systems such as
KEVM~\cite{Hildenbrandt2017} use the $\mathbb K$ framework based on
reachability logic. Due to their reliance on semi-automated theorem provers
which require substantial manual intervention for proof construction, these
formal verification approaches do not scale for analysing the millions of
smart contracts currently deployed on the blockchain.

In contrast to formal verification work for smart contracts, \emph{abstract
interpretation} approaches~\cite{kalra2018,mavridou18} do not require human
intervention; however, they introduce false-positives. The \textsc{zeus}
framework~\cite{kalra2018} translates Solidity source code to LLVM~\cite{LLVM}
before performing the actual analysis in the SeaHorn verification
framework~\cite{SeaHorn}. An alternative approach is that
of~\cite{mavridou18}, in which Solidity code is abstracted to finite-state
automata. 
\footnotetext{Soundy: \textit{"as sound as possible without excessively compromising
precision and/or scalability"} ~\cite{Livshits2015}}

The approach of our Vandal framework~\cite{grech2018} is also partly that of
abstract interpretation.  However, in contrast to the aforementioned abstract
interpretation frameworks, Vandal performs analysis directly on EVM bytecode
using a purpose-built decompiler that translates EVM bytecode to an analyzable
intermediate representation.

\paragraph{Coverage of Vulnerabilities by Existing Tools.}
Oyente~\cite{Luu2016}  identifies four vulnerabilities:
transaction-ordering dependence, timestamp dependence,
exceeding the call stack limit of 1024 (callstack attack) and reentrancy.  
The formal verification tool by~\cite{Bhargavan2016} detects three classes of
vulnerabilities, two of which were covered by Oyente. These include
checking the return value of external address calls, and reentrancy. However,
an upper bound analysis on gas required for a given transaction was created.
These patterns were verified in F* by translating the contracts into F* code,
from which patterns were applied to detect vulnerabilities.  
Similarly, the \textsc{FSolidM} framework~\cite{mavridou18} checks for reentrancy and
transaction ordering vulnerabilities.  It can also detect coding patterns such
as time constraint and authorization issues~\cite{Bartoletti2017}.
The \textsc{maian} framework~\cite{Nikolic2018} focuses
on finding vulnerabilities in smart contracts such as locking of funds
indefinitely, leaking funds to arbitrary users, and smart contracts that can
be killed by anyone.  
The \textsc{gasper}~\cite{chen2017} tool identifies gas-costly patterns in
contract bytecode, often caused by inefficiencies in the Solidity compiler.
The \textsc{zeus} system~\cite{kalra2018} conducts policy checking for a set of
policies including reentrancy, unchecked send, failed send, integer overflow,
transaction state dependence/order and block state dependence. 
teEther has support for detecting vulnerabilities that allow an attacker to
take control of funds, execute third party code, or kill the contract.
Another work focuses exclusively on detecting non-callback free
contracts.~\cite{Grossman2017}

\section{Conclusion}\label{sec:conclusion}

We presented Vandal, a new static analysis framework for detecting security
vulnerabilities in smart contract bytecode. Vandal consists of an analysis
pipeline, including a decompiler that performs abstract interpretation to
translate bytecode to a higher level intermediate representation in the form
of logic relations.  Vandal uses a novel logic-driven approach for defining
security vulnerability analyses, and includes a static analysis library to
ease the development of new analysis specifications. Through a use-case study,
we demonstrated the ease with which vulnerability analyses can be implemented
in Vandal, often requiring only a few lines of \souffle code. Finally, we
performed a large-scale empirical experiment by running Vandal on all 191k
unique smart contracts scraped from the Ethereum blockchain. We showed that
Vandal outperformed the Oyente, EthIR, Mythril, and Rattle analysis tools in
terms of average analysis time and error rate.

\bibliographystyle{WileyNJD-AMA}
\bibliography{references}%

\end{document}